\documentclass[journal]{IEEEtran}
\usepackage{amsmath,amsfonts}
\usepackage{array}
\usepackage{textcomp}
\usepackage{stfloats}
\usepackage{mdwmath}
\usepackage{mdwtab}
\usepackage{epsfig}
\usepackage{eqparbox}
\usepackage{url}
\usepackage{verbatim}
\usepackage{graphicx}
\usepackage{cite}
\usepackage{CJK}
\usepackage{pdfpages}
\usepackage{subfigure} 
\usepackage{float} 
\usepackage{verbatim}
\usepackage{tabularx}
\usepackage{overpic} 
\usepackage{booktabs} 
\usepackage{multirow}
\usepackage{mdwmath}
\usepackage{mdwtab}
\usepackage{diagbox}
\usepackage{amsmath}
\usepackage{amssymb}
\usepackage{siunitx}
\usepackage{mathrsfs}
\usepackage{threeparttable}
\usepackage[ruled,linesnumbered]{algorithm2e}
\usepackage[ruled,linesnumbered]{algorithm2e}
\newcommand{\method}[1]{\texttt{#1}}

\usepackage[colorlinks=true,      
linkcolor=black,      
citecolor=black,      
filecolor=black,      
urlcolor=blue]{hyperref}

\newtheorem{definition}{Definition}

\newtheorem{remark}{Remark}

\newtheorem{lemma}{Lemma}

\begin{document}
	\title{Mixed Platoon Control under Noise and Attacks: Robust Data-Driven Predictive Control and Human-in-the-Loop Validation}
	\author{Shuai Li, Chaoyi Chen, Haotian Zheng, Jiawei Wang, Qing Xu, Jianqiang Wang, and Keqiang Li
		\thanks{This work was supported by the National Natural Science Foundation of China under Grant 52221005, the National Natural Science Foundation of China under Grant 52302410, the China Postdoctoral Science Foundation under Grant 2024T170489, the Postdoctoral Fellowship Program of CPSF under Grant GZB20230354, and the Shuimu Tsinghua Scholarship. Corresponding author: Keqiang Li and Jiawei Wang.}
		\thanks{S.~Li, H.~Zheng, C.~Chen, Q.~Xu, J.~Wang, and K.~Li are with the School of Vehicle and Mobility, Tsinghua University, Beijing, China. (\{li-s21, zhenght24\}@mails.tsinghua.edu.cn, \{chency2023, qingxu, wjqlws, likq\}@tsinghua.edu.cn).}
		\thanks{J.~Wang is with the Department of Civil and Environmental Engineering, University of Michigan, Ann Arbor, USA. (jiawe@umich.edu).}%
	}
	\markboth{}%
	{}

	\maketitle
	
	\begin{abstract}
         Controlling mixed platoons, which consist of both connected and automated vehicles (CAVs) and human-driven vehicles (HDVs), poses significant challenges due to the uncertain and unknown human driving behaviors. Data-driven control methods offer promising solutions by leveraging available trajectory data, but their performance can be compromised by noise and attacks. To address this issue, this paper proposes a Robust Data-EnablEd Predictive Leading Cruise Control (\method{RDeeP-LCC}) framework based on data-driven reachability analysis. The framework over-approximates system dynamics under noise and attack using a matrix zonotope set derived from data, and develops a stabilizing feedback control law. By decoupling the mixed platoon system into nominal and error components, we employ data-driven reachability sets to recursively compute error reachable sets that account for noise and attacks, and obtain tightened safety constraints of the nominal system. This leads to a robust data-driven predictive control framework, solved in a tube-based control manner. Human-in-the-loop experiments demonstrate that the \method{RDeeP-LCC} method significantly improves robustness against noise and attacks, while enhancing tracking accuracy, control efficiency, energy economy, driving comfort, and driving safety.
	\end{abstract}


	\section{Introduction}
	\IEEEPARstart{R}{ecent} advancements in connected and automated vehicles (CAVs) have led to the increasing deployment of vehicles featuring various levels of autonomous driving capabilities in public transportation systems~\cite{wang2018review,alessadrini2023innovative}. Among these innovations, adaptive cruise control (ACC) has emerged as a significant implementation, reducing the need for constant driver intervention in speed management~\cite{de2014effects} and enhancing proactive driving safety~\cite{luo2015intelligent}. Despite these benefits, recent empirical and experimental studies~\cite{gunter2020commercially,makridis2020empirical} have revealed the inherent limitations of ACC in optimizing traffic flow. These limitations are primarily attributed to ACC's over-conservative car-following policy and short-sighted perception capabilities~\cite{shen2023data}.

    In contrast to ACC, cooperative adaptive cruise control (CACC) employs vehicle-to-vehicle (V2V) communication to organize multiple CAVs as a pure CAV platoon and apply cooperative control methods~\cite{boo2023integral}. This approach shows substantial potential in improving traffic performance, including traffic stability~\cite{oncu2014cooperative}, road capacity~\cite{smith2020improving}, and energy efficiency~\cite{ma2019predictive}. However, the effectiveness of CACC is hindered by the requirement for all the involved vehicles to possess autonomous capabilities. In practice, mixed traffic environments, where CAVs and human-driven vehicles (HDVs) coexist, are expected to persist for an extended period~\cite{tajalli2021traffic}. At low CAV penetration rates, the probability of consecutive vehicles being equipped with CAV technology becomes negligible~\cite{hajdu2019robust,jiang2023platoon}. To overcome the limitations of pure CAV platoons, mixed platooning has emerged as a promising alternative, which integrates both CAVs and HDVs in a  platoon~\cite{jin2016optimal,yang2022eco,chen2021mixed}. The core idea is to guide the behavior of HDVs by directly controlling CAVs, thereby enhancing overall traffic performance~\cite{stern2018dissipation,wang2020controllability}. Recent studies, including traffic simulations~\cite{zhan2022data}, hardware-in-the-loop tests~\cite{wang2023implementation}, and real-world experiments~\cite{jin2018experimental}, have shown the potential benefits of mixed platoons for smoothing traffic flow and improving traffic efficiency, even at low CAV penetration rates.

To achieve these benefits while preserving CAV safety, existing research on mixed platoon control mainly relies on model-based control methods. These methods utilize microscopic car-following models, such as the intelligent driver model (IDM)~\cite{treiber2000congested} and optimal velocity model (OVM)~\cite{bando1995dynamical}, to capture the longitudinal behavior of HDVs. Parametric models are then derived to represent the dynamics of the entire mixed platoon system, enabling the implementation of various model-based control strategies, including linear quadratic regulator~\cite{jin2016optimal}, structured optimal control~\cite{wang2020controllability}, model predictive control (MPC)~\cite{feng2021robust}, adaptive control~\cite{liu2023anti}, $\mathcal{H}_\infty$ robust control~\cite{wang2022robustness}, and control barrier function~\cite{zhao2023safety}. However, the inherent randomness and uncertainty in the car-following behavior of HDVs present a significant challenge in accurately identifying the mixed platoon dynamics. The resulting model mismatches may limit the performance of these model-based techniques. 
On the other hand, model-free or data-driven methods have gained increasing attention~\cite{vinitsky2018lagrangian,lan2023safe}. Approaches like adaptive dynamic programming~\cite{huang2020learning} and reinforcement learning~\cite{vinitsky2018lagrangian} have shown potential in learning CAV control policies through iterative training, without the necessity of a previous knowledge about the dynamics of mixed platoons. However, it is worth noting that safety is always prioritized first for CAVs, but these methods lack principled safety constraints, as they typically take an indirect manner by penalizing unsafe actions in the reward function.

For deriving optimal control inputs directly from data, one promising approach is data-driven predictive control, with the combination of the well-established MPC and data-driven techniques~\cite{hewing2020learning}. Along this direction, several methods have been proposed for mixed platoon control~\cite{lan2021data,wang2023deep,wu2023driver,lyu2024kooplcc}, with a notable example being Data-EnablEd Predictive Control (DeePC)~\cite{coulson2019data}. Specifically, DeePC represents the system behavior in a data-centric manner via Willems' fundamental lemma~\cite{willems2005note}, and incorporates explicit input-output constraints in online predictive control optimization. By adapting DeePC to a Leading Cruise Control (LCC) framework~\cite{wang2021leading}, which is particularly designed for mixed traffic, the recently proposed Data-EnablEd Predictive Leading Cruise Control (\method{DeeP-LCC}) allows for CAVs' safe and optimal control in mixed platoons~\cite{wang2023deep}. The effectiveness of this approach has been validated across multiple dimensions, including the mitigation of traffic waves~\cite{wang2023deep}, the reduction of energy consumption~\cite{li2024physics}, and the enhancement of privacy protection~\cite{zhang2023privacy}. However, real-world data are always corrupted by noise from vehicle perception systems or V2X communication channels. Moreover, these data-driven CAV control systems become increasingly vulnerable to attacks in the V2X network, which may maliciously alter control inputs or perceived data to execute attacks~\cite{xu2022reachability}, thereby compromising CAV control safety. Existing research tends to overlook the influence of noise on data collection and online predictive control, and often assumes the absence of adversarial attacks. This assumption could limit the CAV's ability to effectively follow desired trajectories and may raise significant safety concerns~\cite{zhao2024robust}.

To explicitly address noise and attacks, growing evidence has indicated that robustness is crucial in standard DeePC~\cite{huang2023robust,berberich2020data}. Indeed, a recent paper has reformulated \method{DeeP-LCC} using min-max robust optimization to handle unknown disturbances~\cite{shang2024smoothing}. However, prior assumptions on disturbances are still needed to improve computational efficiency. Compared to the min-max approach, reachability analysis offers a more computationally reliable method for ensuring robustness against a wide range of noise and attacks. Several recent works have applied similar techniques to design robust control strategies for CAVs, including anti-attack control employing reach-avoid specification~\cite{xu2022reachability} and formal safety net control using backward reachability analysis~\cite{schurmann2021formal}. Note that most of these methods are model-based, with one notable exception of~\cite{lan2021data}, which presents a zonotopic predictive control (ZPC) approach. Nonetheless, the prediction accuracy in~\cite{lan2021data} may be limited by the utilization of over-approximated data-driven dynamics, which could significantly affect the performance of data-driven predictive control.

To address the aforementioned research gaps, this paper proposes a Robust Data-EnablEd Predictive Leading Cruise Control (\method{RDeeP-LCC}) method that leverages data-driven reachability analysis. The goal is to develop robust control strategies for CAVs against noise and attacks in mixed platoons. We conduct human-in-the-loop experiments to provide near-real-world validation. Additionally, specific evaluation indices are introduced to comprehensively analyze \method{RDeeP-LCC}'s tracking performance under different attack conditions. Some preliminary results have been outlined in~\cite{li2024robust}. Precisely, the main contributions of this paper are as follows:
	\begin{itemize}
		\item[1)]
        We propose a \method{RDeeP-LCC} formulation for mixed platoon control that explicitly addresses noise and attacks. Unlike the existing \method{DeeP-LCC} frameworks~\cite{wang2023deep, li2024physics, shang2024smoothing}, which focus on data-driven predictive control under nominal conditions and remain vulnerable to uncertainties, our approach integrates data-driven reachability analysis into the \method{DeeP-LCC} architecture for the first time. By employing a tube-based control strategy, we systematically design the control inputs of CAVs to ensure that system states remain strictly within prescribed safety constraints under uncertainty. This framework establishes formal safety guarantees, strengthens robustness against noise and attacks, and ensures accurate trajectory tracking in practical applications.
	\end{itemize}

	\begin{itemize}
		\item[2)]
        We conduct comprehensive human-in-the-loop experiments with real human drivers using driving simulators. To the best of our knowledge, this is the first human-in-the-loop validation of data-driven predictive control methods in mixed platoons. Experimental results show that without explicit consideration of noise and attacks, existing methods fail to provide practical safety guarantees. By contrast, \method{RDeeP-LCC} outperforms the baselines by enhancing driving safety by $93.5\%$, while also achieving system-wide improvements: reducing velocity deviations by $26.1\%$, real cost by $24.7\%$, fuel consumption by $11.4\%$, and improving driving comfort by $26.0\%$. These findings highlight the practical effectiveness of \method{RDeeP-LCC} in controlling mixed platoons under challenging operating conditions.
	\end{itemize}
    
       The remainder of this paper is organized as follows: Section~\ref{Sec:2} introduces the problem statement and preliminaries. Section~\ref{Sec:3} presents the \method{RDeeP-LCC} formulation. Human-in-the-loop experiments are provided in Section~\ref{Sec:5}, and Section~\ref{Sec:6} concludes this paper.

	\section{Problem Statement}
	\label{Sec:2}
    In this section, we first introduce the research scenario, and then give the parametric model of the mixed platoon system under the LCC framework~\cite{wang2021leading}.
    
	\subsection{Research Scenario}
\begin{figure}[t]
    	\centering
    	\includegraphics[width=8.8cm]{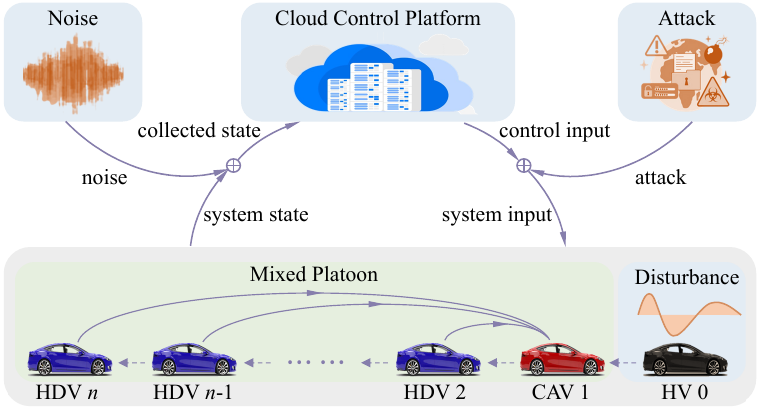}\\
    	\vspace{-0.25cm}
    	\caption{Schematic for the mixed platoon under the influence of noise and attacks. The noise and attacks affect the uplink and downlink of the cloud control platform, respectively.} 
    	\label{Fig:MixedPlatoon}
             \vspace{-0.2cm}
    \end{figure}

We consider a typical mixed platoon system with a Leading Cruise Control (LCC) topology~\cite{wang2021leading}, as illustrated in~\figurename~\ref{Fig:MixedPlatoon}. Each mixed platoon consists of one leading CAV (indexed as $1$) and $n-1$ following HDVs (indexed as $2,\ldots,n$), with all vehicles following a head vehicle (HV) (indexed as $0$).  In this configuration, the CAV plays a pivotal role in regulating the motion of the entire platoon by responding to the motion of the HV.

To enable coordinated control, we adopt an edge cloud control framework. Specifically, roadside units are deployed to collect real-time data from all vehicles and transmit it to the cloud control platform. Then, a specific controller within the cloud calculates the control inputs, which are sent to the CAV to regulate mixed platoons. In this setting, even small velocity changes in the HV must be synchronized across all following vehicles to maintain safety. Given the fact that only the CAV is under direct control, the dynamic process of velocity adaptation poses substantial challenges, particularly in scenarios where the system is subject to noise and attacks. Accordingly, the main research focus of this paper is to develop a robust data-driven control method for CAV that can effectively mitigate the impact of noise and attacks.

 \begin{remark}
      For large-scale mixed traffic, our method can be extended in a decentralized manner based on spatial segmentation~\cite{li2025influence}. Specifically, the traffic flow is partitioned into multiple mixed platoons, where each platoon consists of one CAV and $n-1$ following HDVs. This partition is naturally determined by the spatial locations of the CAVs in mixed traffic. Then, the proposed method decomposition transforms the original high-dimensional, multi-input system into multiple lower-dimensional, single-input subsystems, significantly reducing system complexity. Consequently, this facilitates the efficient design and real-time implementation of the control algorithm. Moreover, the distributed nature of this strategy ensures inherent scalability, enabling it to adapt effectively to traffic systems with varying sizes.
\end{remark}

    \subsection{Parametric Model of Mixed Platoon System}
	In the following, we provide a concise overview of the parametric modeling process for mixed platoons. The system state is defined as $x(t)= \left[x_1^{\top}(t),x_2^{\top}(t),\ldots,x_n^{\top}(t)\right]^{\top} \in \mathbb{R}^{2n \times 1}$, where $x_i(t)=\left[\tilde{s}_i(t), \tilde{v}_i(t)\right]^{\top}$ consists of the spacing error $\tilde{s}_i(t)$ and velocity error $\tilde{v}_i(t)$ of the $i$-th vehicle. Then, the continuous-time model of the mixed platoon is given by:
	\begin{equation}
	\label{Eq:ContinuousSystem}
	\dot{x}(t)=A_{\rm{con}}x(t)+B_{\rm{con}}u(t)+H_{\rm{con}}\epsilon(t) +J_{\rm{con}}\vartheta(t),
	\end{equation}	  
    where $u(t) = u_1(t)\in \mathbb{R}$ and $\vartheta(t) = \vartheta_1(t)\in \mathbb{R}$ represents the control input and attack input for the CAV, $\epsilon(t)=\tilde{v}_0(t) \in \mathbb{R}$ is the disturbance from the HV. The matrices $A_{\rm{con}} \in \mathbb{R}^{2n \times 2n}$, $B_{\rm{con}} \in \mathbb{R}^{2n \times 1}$, $H_{\rm{con}} \in \mathbb{R}^{2n \times 1}$, and $J_{\rm{con}}\in \mathbb{R}^{2n \times 1} $ denote the system, control, disturbance, and attack matrices. A detailed derivation of \eqref{Eq:ContinuousSystem} can be found in~\cite{wang2023deep,wang2020controllability}.

	Subsequently, the continuous-time model~\eqref{Eq:ContinuousSystem} is discretized using the forward Euler method, with noise $\omega(k)$ included, resulting in the discrete-time model:
	\begin{equation}
	\label{Eq:DiscreteSystem}
	{x}(k+1)=Ax(k)+Bu(k)+H \epsilon (k)+ J\vartheta(k)+\omega(k),
	\end{equation}
	where $k$ represents the time step, $A$, $B$, $H$, and $J$ are the corresponding discrete-time matrices. Note that the model~\eqref{Eq:DiscreteSystem} is indeed unknown, and is only used to clarify the dimensions and physical meaning of state and control variables.
    
    Motivated by the existing research~\cite{xu2022reachability,fan2018controller,khoshnevisan2025secure,sun2023secure}, we assume that the disturbance, attack, and noise are bounded:
    \begin{equation}
		\label{Eq:W_Bound}
		\left \|  \epsilon(k) \right \| _{\infty } \leq \epsilon_\mathrm{max},\quad 
            \left \|  \vartheta(k) \right \| _{\infty } \leq \vartheta_\mathrm{max},\quad 
            \left \|  \omega(k) \right \| _{\infty } \leq \omega_\mathrm{max},
  \end{equation}
where $\epsilon_\mathrm{max}$, $\vartheta_\mathrm{max}$, and $\omega_\mathrm{max}$ denote the respective upper bounds. Note that these bounds must be known a priori, as they are essential for the offline construction of the over-approximated system matrix set~\eqref{Eq:M_ABHJ} and the online computation of the data-driven reachable set~\eqref{Eq:ReachableSet}, as detailed in Section~\ref{Sec:3}.

\begin{remark}
    In this paper, we do not explicitly identify the model of FDI attacks, as such attacks are often unpredictable and difficult to characterize~\cite{sun2023secure,song2025model}. Instead, we consider a general class of attacks where the injected signal $\vartheta(k)$ is bounded and may be either state-dependent or state-independent, without assuming any specific structure or distribution. This general formulation enhances the applicability of the proposed method to a broader range of adversarial scenarios.
\end{remark}

 \subsection{Preliminaries and Theoretical Foundations}
    \label{Sec:2C}
    
Before proceeding, we introduce the necessary preliminaries on data-driven control and reachable sets, which form the foundation of our proposed approach. The data-driven control method employed in this paper is based on Willems' fundamental lemma and Hankel matrices, with relevant definitions provided in~\cite[Theorem 1]{willems2005note} and~\cite[Definition 4.4]{coulson2019data}. For the reachable sets, we utilize zonotope sets to describe the reachable sets for efficient computation~\cite{alanwar2023data}. Some basic definitions are presented below. 
	\begin{definition}[Interval Set~\cite{althoff2010reachability}]
		\label{Definition:IntervalSet}
		An interval set $ \mathcal{I}$ is a connected subset of $\mathbb{R} ^{n}$, and it can be defined as $ \mathcal{I}=\left\{x_\mathcal{I} \in \mathbb{R}^{n} \mid \underline{x}_{\mathcal{I}_i} \leq x_{\mathcal{I}_i} \leq \overline{x}_{\mathcal{I}_i} \quad \forall i=1, \ldots, n\right\}$, where $ \underline{x}_{\mathcal{I}_i} $ and $\overline{x}_{\mathcal{I}_i}$ are the lower bound and upper bound of $x_{\mathcal{I}_i}$, respectively. Interval set can be represented as $\mathcal{I}=[\underline{\mathcal{I}},\overline{\mathcal{I}}]$, with $\underline{\mathcal{I}} =[\underline{x}_{\mathcal{I}_1},\underline{x}_{\mathcal{I}_2},\ldots,\underline{x}_{\mathcal{I}_n}]$
		and $\overline{\mathcal{I}}= [\overline{x}_{\mathcal{I}_1},\overline{x}_{\mathcal{I}_2},\ldots,\overline{x}_{\mathcal{I}_n}]$.
	\end{definition}

         \vspace{0.1cm}
	\begin{definition}[Zonotope Set~\cite{kuhn1998rigorously}]
	\label{Definition:ZonotopeSet}
		 Given a center vector $c_\mathcal{Z} \in \mathbb{R} ^n$, and $\gamma_\mathcal{Z} \in \mathbb{N}$ generator vectors in a generator matrix $G_\mathcal{Z} =\left[g_\mathcal{Z}^{(1)}, g_\mathcal{Z}^{(2)},\ldots, g_\mathcal{Z}^{(\gamma_\mathcal{Z})}\right] \in \mathbb{R}^{n \times \gamma_\mathcal{Z}}$, a zonotope set is defined as $\mathcal{Z} =\left \langle c_\mathcal{Z}, G_\mathcal{Z}\right \rangle = \left\{x \in \mathbb{R}^n \mid x=c_\mathcal{Z}+\sum_{i=1}^{\gamma_\mathcal{Z}} \beta^{(i)} g_\mathcal{Z}^{(i)},-1 \leq \beta^{(i)} \leq 1\right\}$.
	For zonotope sets, the following operations hold:
	
\begin{itemize}
	\item  \textit{Linear Map:} For a zonotope set $\mathcal{Z}=\left \langle c_\mathcal{Z}, G_\mathcal{Z}\right \rangle $, $L \in \mathbb{R}^{m\times n}$, the linear map is defined as $L\mathcal{Z}=\left \langle Lc_\mathcal{Z}, LG_\mathcal{Z}\right \rangle$.
	
	\item  \textit{Minkowski Sum:} Given two zonotope sets $\mathcal{Z}_1=\left \langle c_{\mathcal{Z}_1}, G_{\mathcal{Z}_1}\right \rangle$ and $\mathcal{Z}_2=\left \langle c_{\mathcal{Z}_2}, G_{\mathcal{Z}_2}\right \rangle$ with compatible dimensions, the Minkowski sum is defined as $\mathcal{Z}_1 + \mathcal{Z}_2=\left\langle c_{\mathcal{Z}_1}+c_{\mathcal{Z}_2},\left[G_{\mathcal{Z}_1}, G_{\mathcal{Z}_2}\right]\right\rangle$. 
	
	\item  \textit{Cartesian Product:} Given two zonotope sets $\mathcal{Z}_1=\left \langle c_{\mathcal{Z}_1}, G_{\mathcal{Z}_1}\right \rangle$ and $\mathcal{Z}_2=\left \langle c_{\mathcal{Z}_2}, G_{\mathcal{Z}_2}\right \rangle$, the cartesian product is defined as
	\vspace{-0.1cm}
	\begin{equation*}
	\label{Cartesian}
	\mathcal{Z}_1 \times\mathcal{Z}_2  =\left\langle\begin{bmatrix}
	c_{\mathcal{Z}_1} \\
	c_{\mathcal{Z}_2}
	\end{bmatrix},\begin{bmatrix}
	G_{\mathcal{Z}_1} & 0 \\
	0 & G_{\mathcal{Z}_2}
	\end{bmatrix}\right\rangle.
	\end{equation*}
       
       \item  \textit{Over-Approximated Using Interval Set:} A zonotope set $\mathcal{Z}=\left \langle c_\mathcal{Z}, G_\mathcal{Z}\right \rangle$ could be over-approximated by an interval set $\mathcal{I}=[c_\mathcal{Z}-\bigtriangleup {g_\mathcal{Z}},c_\mathcal{Z}+\bigtriangleup {g_\mathcal{Z}}]$, where $\bigtriangleup {g_\mathcal{Z}} = \sum_{i=1}^{\gamma_\mathcal{Z}}\left|g_\mathcal{Z}^{(i)}\right|$.
    \end{itemize}
   \end{definition}

	\begin{definition}[Matrix Zonotope Set~\cite{althoff2010reachability}]
	\label{Definition:MatrixZonotopeSet}
		Given a center matrix $C_\mathcal{M} \in \mathbb{R} ^{n \times m}$, and $\gamma_\mathcal{M} \in \mathbb{N}$ generator matrices in a generator matrix $G_\mathcal{M} =\left[G_\mathcal{M}^{(1)}, G_\mathcal{M}^{(2)},\ldots, G_\mathcal{M}^{(\gamma_\mathcal{M})}\right] \in \mathbb{R}^{n \times m\gamma_\mathcal{M}}$, a matrix zonotope set is defined as $\mathcal{M} =\left \langle C_\mathcal{M}, G_\mathcal{M}\right \rangle = \left\{X \in \mathbb{R}^{n \times m} \mid X=C_\mathcal{M}+\sum_{i=1}^{\gamma_\mathcal{M}} \beta^{(i)} G_\mathcal{M}^{(i)},-1 \leq \beta^{(i)} \leq 1\right\}$.
	\end{definition}

	\section{Methodology}
	\label{Sec:3}
 
\begin{figure*}[htbp]
	\centering
	{\includegraphics[width=18cm]{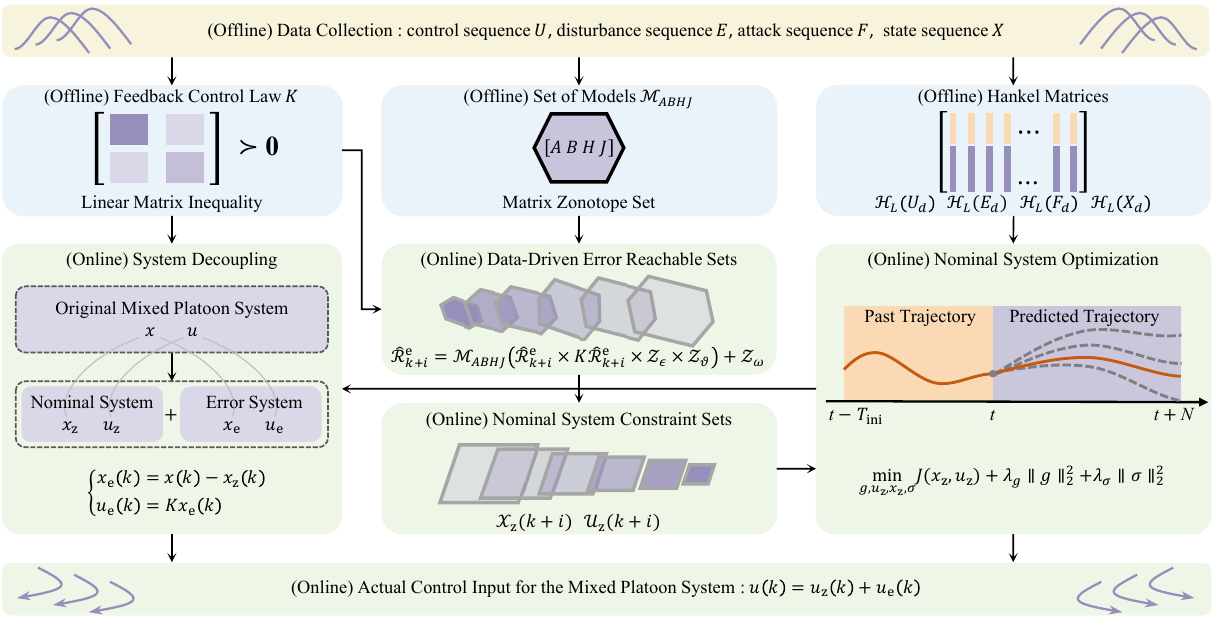}}\\
	\vspace{-0.25cm}
	\caption{Schematic of the proposed \method{RDeeP-LCC} method for mixed platoons. In the offline learning phase (blue), the method utilizes collected data (yellow) to calculate the over-approximated system matrix set $\mathcal{M}_{\scalebox{0.7}{ABHJ}}$, derive a data-driven feedback control law $K$ to ensure stability for all possible systems, and generate the Hankel matrices. In the online control phase (green), the \method{RDeeP-LCC} solves for optimal control input for the CAV in a receding horizon strategy. Specifically, the system is decomposed into the error system and the nominal system. Using $\mathcal{M}_{\scalebox{0.7}{ABHJ}}$ and $K$, the method recursively derives the data-driven reachable set of error states, and then subtracts this set from the constraints of the original system to obtain a more compact nominal system constraint. Then, the nominal control input $u_\mathrm{z}(k)$ is calculated using the standard \method{DeeP-LCC} under the compact nominal constraint. Finally, the actual control input of the CAV is obtained by combining the nominal control input $u_\mathrm{z}(k)$ with the error feedback control input $u_\mathrm{e}(k)$ in a tube-based control manner.}
	\label{Fig:RDeeP}
    \end{figure*}

    This section proposes the Robust Data-EnablEd Predictive Leading Cruise Control (\method{RDeeP-LCC}) method for mixed platoon control, including data collection, offline learning, and online control, as shown in~\figurename~\ref{Fig:RDeeP}.

    \subsection{Data Collection}
    \label{Sec:3A}
    In this study, we applies a sequence of persistently exciting inputs $u(k)$, $\epsilon(k)$, and $\vartheta(k)$ with a length $ T+1 $ to the mixed platoon system for data collection. Specifically, the control input sequence $U_-$, the disturbance input sequence $E_-$, the attack input sequence $F_-$, and the corresponding state sequence $X_-$ and $X_+$ are defined as follows:
    \begin{subequations}
    \label{Eq:data_sequence}
    	\begin{equation}
            \label{Eq:Sequence_U}
    	U_-=[u(1),u(2),\ldots,u(T)] \in \mathbb{R} ^{1 \times T},
    	\end{equation}
    	\begin{equation}
            \label{Eq:Sequence_E}
    	E_-=[\epsilon(1),\epsilon(2),\ldots,\epsilon(T)] \in \mathbb{R} ^{1 \times T},
    	\end{equation}
    	\begin{equation}
            \label{Eq:Sequence_F}
    	F_-=[\vartheta(1),\vartheta(2),\ldots,\vartheta(T)] \in \mathbb{R} ^{1 \times T},
    	\end{equation}
    	\begin{equation}
            \label{Eq:Sequence_X}
    	X_-=[x(1),x(2),\ldots,x(T)] \in \mathbb{R} ^{2n \times T},
    	\end{equation}
    	\begin{equation}
    	X_+=[x(2),x(3),\ldots,x(T+1)] \in \mathbb{R} ^{2n \times T}.
    	\end{equation}
    \end{subequations}
    In addition, the sequence of unknown noise is denoted as
    \begin{equation}
     \label{Eq:Sequence_W}
	W_-=[\omega(1),\omega(2),\ldots,\omega(T)] \in \mathbb{R} ^{2n \times T},
	\end{equation}    
    although it is important to note that $W_-$ is not measurable. This notation is introduced solely to facilitate the proof of Lemma~\ref{Lemma:M_ABHJ}, and does not imply that $W_-$ is measurable.

 \begin{remark}
      In this study, the noise $\omega(k)$ in~\eqref{Eq:Sequence_W} is assumed to be bounded but not measurable. The disturbance $\epsilon(k)$ in~\eqref{Eq:Sequence_E} is the velocity deviation of the HV from its equilibrium state. Since the velocity of the HV is directly measurable and its equilibrium state can be estimated, $\epsilon(k)$ can be explicitly determined. Finally, the historical attack  $\vartheta(k)$ in~\eqref{Eq:Sequence_F} is also reconstructible. This is because the control input $u(k)$ computed by the cloud control platform is known, and the actual control input $u(k) + \vartheta(k)$, which is essentially the digital signal received by the vehicle control unit, is also known and transmitted back to the cloud control platform. By taking the difference between these two signals, the historical values of $\vartheta(k)$ can be accurately obtained.
\end{remark}

\subsection{Offline Learning}
\label{Sec:3B}
In the offline learning phase, given that the matrices $A$, $B$, $H$, and $J$ in~\eqref{Eq:DiscreteSystem} are unknown, we utilize collected data to construct an over-approximated system matrix set $\mathcal{M}_{\scalebox{0.7}{ABHJ}}$. This set models the unknown and uncertain dynamics of the mixed platoon system. We then derive a data-driven stabilizing feedback control law $K$ from data to ensure stability across all possible systems configurations represented by $\begin{bmatrix}A~B\end{bmatrix}$. Additionally, the Hankel matrices are formed using the collected data and serve as part of the predictor. These components are later integral to the online control phase, discussed in Section~\ref{Sec:3C}.

\subsubsection{ Over-Approximated System Matrix Set} 
We first construct the matrix zonotope set $\mathcal{M}_{\scalebox{0.7}{ABHJ}}$ to over-approximate all possible system models $\begin{bmatrix}A~B~H~J\end{bmatrix}$ that are consistent with the noisy data. The following Lemma \ref{Lemma:M_ABHJ} is needed.

	\begin{lemma}      
		\label{Lemma:M_ABHJ}
		Given the data sequences $U_{-}$, $E_{-}$, $F_{-}$, $X_{-}$, and $X_{+}$ from the mixed platoon system~\eqref{Eq:DiscreteSystem}, and transforming the bounded forms of the disturbance  $\epsilon(k)$, the attack $\vartheta(k)$, and the noise $\omega(k)$ in~\eqref{Eq:W_Bound} to be zonotope sets, given by:		
		\begin{equation}
		\label{Eq:W_Zonotope}
            \begin{cases}
    	\epsilon(k) \in \mathcal{Z}_{\epsilon}=\left \langle c_{\mathcal{Z}_{\epsilon}}, G_{\mathcal{Z}_{\epsilon}}\right \rangle=\left \langle 0, \epsilon_\mathrm{max}\right \rangle,	 \\
    	\vartheta(k)\in \mathcal{Z}_{\vartheta}=\left \langle c_{\mathcal{Z}_{\vartheta}}, G_{\mathcal{Z}_{\vartheta}}\right \rangle=\left \langle 0, \vartheta_\mathrm{max}\right \rangle, \\
    	\omega(k) \in \mathcal{Z}_{\omega}=\left \langle c_{\mathcal{Z}_{\omega}}, G_{\mathcal{Z}_{\omega}}\right \rangle=\left \langle \mathbf{0}_{2n \times 1}, \omega_\mathrm{max} \mathbf{I}_{2n \times 2n}\right \rangle,\\
    	\end{cases}
		\end{equation}
        where $\mathbf{0}$ and $\mathbf{I}$ denote the zero and identity matrices of appropriate dimensions, respectively.
        
		If the matrix $ \begin{bmatrix} X_{-}^{\top}~U_{-}^{\top}~E_{-}^{\top}~F_{-}^{\top}\end{bmatrix}^{\top}$ is of full row rank, then the set of all possible $\begin{bmatrix}A~B~H~J\end{bmatrix}$ can be obtained:
		\begin{equation}
		\label{Eq:M_ABHJ}
		\mathcal{M}_{\scalebox{0.7}{ABHJ}}=\left(X_{+}-\mathcal{M}_{\omega}\right)\begin{bmatrix}
		X_{-} \\
		U_{-} \\
		E_{-} \\
		F_{-} 
		\end{bmatrix}^{\dagger},
		\end{equation}
		where $\dagger$ is the Moore–Penrose pseudoinverse. The term $\mathcal{M}_{\omega}$ represents the noise matrix zonotope set:
  \begin{equation}
  \label{eq:setofnoise}
  \mathcal{M}_{\omega} = \left \langle C_{\mathcal{M}_{\omega}}, \left[G_{\mathcal{M}_{\omega}}^{(1)}, G_{\mathcal{M}_{\omega}}^{(2)},\ldots, G_{\mathcal{M}_{\omega}}^{(2nT)}\right]\right \rangle,
  \end{equation}
  which is derived from the noise zonotope $\mathcal{Z}_{\omega} = \left \langle c_{\mathcal{Z}_{\omega}}, G_{\mathcal{Z}_{\omega}}\right \rangle $ in~\eqref{Eq:W_Zonotope}, with the specific formulations given as
		\begin{subequations}
			\begin{equation}
			C_{\mathcal{M}_{\omega}}=\begin{bmatrix}
			c_{\mathcal{Z}_{\omega}} & \ldots & c_{\mathcal{Z}_{\omega}}
			\end{bmatrix} \in \mathbb{R} ^{2n \times T},
			\end{equation}
			\begin{equation}
			G_{\mathcal{M}_{\omega}}^{(1+(i-1) T)}=\begin{bmatrix}
			g_{\mathcal{Z}_{\omega}}^{(i)} & \mathbf{0}_{2n \times(T-1)}
			\end{bmatrix} \in \mathbb{R} ^{2n \times T},
			\end{equation}
			\begin{equation}
			G_{\mathcal{M}_{\omega}}^{(j+(i-1) T)}=\begin{bmatrix}
			\mathbf{0}_{2n \times(j-1)} & g_{\mathcal{Z}_{\omega}}^{(i)} & \mathbf{0}_{2n \times(T-j)} 
			\end{bmatrix} \in \mathbb{R} ^{2n \times T},
			\end{equation}
			\begin{equation}
			G_{\mathcal{M}_{\omega}}^{(T+(i-1) T)}=\begin{bmatrix}
			\mathbf{0}_{2n \times(T-1)} & g_{\mathcal{Z}_{\omega}}^{(i)}
			\end{bmatrix} \in \mathbb{R} ^{2n \times T},
			\end{equation}
		\end{subequations}
		where $g_{\mathcal{Z}_{\omega}}^{(i)}$ denotes the $i$-th column of $G_{\mathcal{Z}_{\omega}}$ in~\eqref{Eq:W_Zonotope}, with $\forall i=\left\{1, 2,\ldots, 2n\right\}$, and $j=\{2, 3, \ldots, T-1\}$.
	\end{lemma}

    \textit{Proof:} For the system description in~\eqref{Eq:DiscreteSystem}, we have
	\begin{equation}
	X_{+} = \begin{bmatrix}A~B~H~J\end{bmatrix}\begin{bmatrix}
	X_{-} \\
	U_{-} \\
	E_{-} \\
	F_{-} 
	\end{bmatrix} + W_{-}.
	\end{equation}
	Since the matrix $ \begin{bmatrix}X_{-}^{\top}~U_{-}^{\top}~E_{-}^{\top}~F_{-}^{\top}\end{bmatrix}^{\top} $ is of full row rank, then we could get 
	\begin{equation}
	\begin{bmatrix}A~B~H~J\end{bmatrix} =\left(X_{+}-W_{-}\right)\begin{bmatrix}
	X_{-} \\
	U_{-} \\
	E_{-} \\
	F_{-} 
	\end{bmatrix}^{\dagger},
	\end{equation}
	where the noise $W_{-}$ in the collected data is unknown, but one can use the corresponding bounds $\mathcal{M}_{\omega}$ to obtain~\eqref{Eq:M_ABHJ}. Then, the matrix zonotope set $\mathcal{M}_{\scalebox{0.7}{ABHJ}}$ is an over-approximation for system models $\begin{bmatrix}A~B~H~J\end{bmatrix}$ considering noisy data. $\hfill\blacksquare$

\subsubsection{Data-Driven Stabilizing Feedback Control Law}
We then aim to stabilize all possible $\begin{bmatrix}A~B\end{bmatrix}$ by a feedback law $K$. Collect data under the condition that $\epsilon(k)=0$ and $ \vartheta(k)=0$, which are straightforward to achieve. Inspired by~\cite{van2023quadratic}, we assume the data sequence $W_-$ satisfies a quadratic matrix inequality:
	\begin{equation}
	\label{Eq:Assume_HEW}
	\begin{bmatrix}
	I \\
	W_-
	\end{bmatrix}^{\top} \Phi \begin{bmatrix}
	I \\
	W_-
	\end{bmatrix} \geq 0,
	\end{equation}
	where $ \Phi = \begin{bmatrix}
		\Phi_{11} ~ \Phi_{12} \\
		\Phi_{21} ~ \Phi_{22}
	\end{bmatrix} \in \mathbb{S}^{2n+T}$, with $ \Phi_{11} \in \mathbb{S}^{2n}$, $ \Phi_{12} \in \mathbb{R}^{2n\times T}$, $ \Phi_{21} \in \mathbb{R}^{T \times 2n}$, $ \Phi_{22} \in \mathbb{S}^{T}$. Based on the bound of $ \omega(k)$ in $W_-$ as described in~\eqref{Eq:W_Bound}, where $\left \|  \omega(k) \right \| _{\infty } \leq \omega_\mathrm{max}$, we set $ \Phi_{22} = -I $, $ \Phi_{12} = 0 $, and $ \Phi_{11} = \omega_\mathrm{max}^2 TI $.
	
	Then, the feedback control law $K$ that stabilizes all possible systems $\begin{bmatrix}A~B\end{bmatrix}$ can be obtained by Lemma~\ref{Lemma:K}.
	
	\begin{lemma}      
		\label{Lemma:K}
		For mixed platoon systems~\eqref{Eq:DiscreteSystem}, if the assumption in~\eqref{Eq:Assume_HEW} hold and the matrix $ \begin{bmatrix} X_{-}^{\top}~U_{-}^{\top} \end{bmatrix}^{\top}$ is of full row rank, one can solve the following linear matrix inequalities (LMIs):
		\begin{subequations}
		\begin{equation}
		\begin{bmatrix}
		P & 0 \\
		0 & -P
		\end{bmatrix}-\begin{bmatrix}
		I & X_{+} \\
		0 & -X_{-}
		\end{bmatrix} \Phi \begin{bmatrix}
		I & X_{+} \\
		0 & -X_{-}
		\end{bmatrix}^{\top}>0,
		\end{equation}
		\begin{equation}
		P-\begin{bmatrix}
		I & X_{+}\end{bmatrix} \Phi \begin{bmatrix}
		I \\
		X_{+}^{\top}
		\end{bmatrix}+\Theta \begin{bmatrix}
		X_{-} \\
		U_{-}
		\end{bmatrix}^{\top} \Psi \begin{bmatrix}
		X_{-} \\
		U_{-}
		\end{bmatrix} \Theta^{\top}>0,
		\end{equation}
		\begin{equation}
		\Psi = \left(\begin{bmatrix}
		X_{-} \\
		U_{-}
		\end{bmatrix} \Phi_{22} \begin{bmatrix}
		X_{-} \\
		U_{-}
		\end{bmatrix}^{\top}\right)^{-1},
		\end{equation}
		\end{subequations}
		to obtain the positive definite matrix $ P$, where $\Theta = \Phi_{12} + X_{+}\Phi_{22} $. Using $P$, the feedback gain can be obtained by
		\begin{equation}
		\label{Eq:K}
		K=\left(U_{-}\left(\Phi_{22}+\Theta^{\top} \Gamma^{\dagger} \Theta\right) X_{-}^{\top}\right)\left(X_{-}\left(\Phi_{22}+\Theta^{\top} \Gamma^{\dagger} \Theta\right) X_{-}^{\top}\right)^{\dagger},
		\end{equation}
		which stabilizes all possible systems $\begin{bmatrix}A~B\end{bmatrix}$, where 
  \begin{equation} 
  \Gamma = P-\begin{bmatrix}
		I & X_{+}\end{bmatrix} \Phi \begin{bmatrix}
		I \\
		X_{+}^{\top}
		\end{bmatrix}.
  \end{equation}
	\end{lemma}

    \textit{Proof:} Lemma \ref{Lemma:K} is derived from~\cite[Theorem 5.3]{van2023quadratic}, with a detailed proof available in~\cite{van2023quadratic}.$\hfill\blacksquare$

\subsubsection{Hankel Matrices}
    We finally utilize the collected data to form the Hankel matrices. In particular, these matrices are partitioned into two parts, corresponding to the trajectory data in the past $ T_{\mathrm{ini}} \in \mathbb{N}$ steps and in the future $ N \in \mathbb{N}$ steps, defined as follows:
    \begin{equation}
    \label{Eq:Hankel}
    \begin{split}
    &\begin{bmatrix}
    U_{\mathrm{p}} \\
    U_{\mathrm{f}}
    \end{bmatrix}=\mathcal{H}_{L}({\mathrm{col}}(U_-)),
    \begin{bmatrix}
    E_{\mathrm{p}} \\
    E_{\mathrm{f}}
    \end{bmatrix}=\mathcal{H}_{L}({\mathrm{col}}(E_-)),\\
    &\begin{bmatrix}
    F_{\mathrm{p}} \\
    F_{\mathrm{f}}
    \end{bmatrix}=\mathcal{H}_{L}({\mathrm{col}}(F_-)),
    \begin{bmatrix}
    X_{\mathrm{p}} \\
    X_{\mathrm{f}}
    \end{bmatrix}=\mathcal{H}_{L}({\mathrm{col}}(X_-)),
    \end{split}
    \end{equation}
    where $L=T_{\mathrm{ini}}+N$, and $U_{\mathrm{p}} , U_{\mathrm{f}} $ contain the upper $T_{\mathrm{ini}}$ rows and lower $ N $ rows of $\mathcal{H}_{L}({\mathrm{col}}(U_-))$, respectively (similarly for $E_{\mathrm{p}}$ and $E_{\mathrm{f}}$, $F_{\mathrm{p}}$ and $F_{\mathrm{f}}$, $X_{\mathrm{p}}$ and $X_{\mathrm{f}}$).

    \subsection{Online Control}
    \label{Sec:3C}
    To ensure the robustness of the mixed platoon control system under noise and attacks, the reachable set technique is introduced. Inspired by the tube-based control method~\cite{mayne2005robust} and the reach-avoid control method~\cite{fan2018controller}, the system is first decoupled into a nominal system and an error system. The reachable set of the error system is computed under disturbances, attacks, and noise. Then, the reachable set of the nominal system is obtained by subtracting the error reachable set from the practical constraints. Using these compact nominal constraints, the nominal control input is computed via the standard \method{DeeP-LCC}. Finally, the actual control input for the CAV is obtained by combining the nominal control input with the error feedback control input.
    
    \subsubsection{Mixed Platoon System Decoupling}
    We start by employing the parametric system model~\eqref{Eq:DiscreteSystem} to illustrate the decoupling process. The decoupled nominal system and the error system are denoted as follows:
	  \begin{subequations}
		\label{Eq:DynamicsDecomposition}
		\begin{equation}
		\label{Eq:NominalDynamics}
		{x}_\mathrm{z}(k+1)=A{x}_\mathrm{z}(k)+B{u}_\mathrm{z}(k)+H\epsilon_\mathrm{z}(k) + J\vartheta_\mathrm{z}(k) ,
		\end{equation}
		\begin{equation}
		\label{Eq:ErrorDynamics}
		{x}_\mathrm{e}(k+1)=A{x}_\mathrm{e}(k)+Bu_\mathrm{e}(k)+H\epsilon_\mathrm{e}(k)+ J\vartheta_\mathrm{e}(k)+ \omega(k),
		\end{equation}
	\end{subequations}
	where ${x}_\mathrm{z}(k)$, $u_\mathrm{z}(k)$, $\epsilon_\mathrm{z}(k)$, $\vartheta_\mathrm{z}(k)$ and ${x}_\mathrm{e}(k)$, $u_\mathrm{e}(k)$, $\epsilon_\mathrm{e}(k)$, $\vartheta_\mathrm{e}(k)$ represent the state, control input, disturbance input, and attack input of the nominal dynamics system and error dynamics system, respectively. Specifically, we have
	\begin{equation}
	\label{Eq:DecoupledSystem}
	\begin{cases}
	{x}(k) ={x}_\mathrm{z}(k) + {x}_\mathrm{e}(k),	 \\
	{u}(k) ={u}_\mathrm{z}(k) + {u}_\mathrm{e}(k), \\
	{\epsilon}(k) ={\epsilon}_\mathrm{z}(k) + {\epsilon}_\mathrm{e}(k),\\
	{\vartheta}(k) ={\vartheta}_\mathrm{z}(k) + {\vartheta}_\mathrm{e}(k).\\
	\end{cases}
	\end{equation}
	Particularly, we set
	\begin{equation}
	\label{Eq:DisturbanceDecouple}
	{\epsilon}_\mathrm{z}(k) = 0 ,\quad {\epsilon}_\mathrm{e}(k) = {\epsilon}(k),\quad {\vartheta}_\mathrm{z}(k) = 0 ,\quad {\vartheta}_\mathrm{e}(k) = {\vartheta}(k),
	\end{equation}
	so that {the disturbance $\epsilon(k)$,  the attack $\vartheta(k)$, and the noise $\omega(k)$} are considered exclusively in the error system, without affecting the nominal system. This simplifies the solution of the \method{RDeeP-LCC} optimization formulation in the following.

\subsubsection{Data-Driven Reachable Set of Error State}
     Based on the noise, disturbance, and attack zonotope sets defined in~\eqref{Eq:W_Zonotope}, the model matrix zonotope set $\mathcal{M}_{\scalebox{0.7}{ABHJ}}$ derived in~\eqref{Eq:M_ABHJ}, and linear state feedback gain $K$ derived in~\eqref{Eq:K},  we can compute the error state reachable set using Lemma~\ref{Lemma:ReachableSet} for the error dynamics described in~\eqref{Eq:ErrorDynamics}.

	\begin{lemma}      
		\label{Lemma:ReachableSet}
    For the system described by~\eqref{Eq:ErrorDynamics}, given input-state trajectories $U_{-}$, $E_{-}$, $F_{-}$, $X_{-}$, $X_{+}$, if the matrix $ \begin{bmatrix}X_{-}^{\top} ~ U_{-}^{\top} ~ E_{-}^{\top} ~ F_{-}^{\top}\end{bmatrix}^{\top} $ is of full row rank, then the recursive relation for the data-driven error state reachable set can be computed as follows:
		\begin{equation}
		\label{Eq:ReachableSet}
		\hat{\mathcal{R}}^\mathrm{e}_{k+i+1}=\mathcal{M}_{\scalebox{0.7}{ABHJ}}\left(\hat{\mathcal{R}}^\mathrm{e}_{k+i} \times K\hat{\mathcal{R}}^\mathrm{e}_{k+i} \times \mathcal{Z}_{\epsilon} \times \mathcal{Z}_{\vartheta}\right)+\mathcal{Z}_{\omega},
		\end{equation}
where $\hat{\mathcal{R}}^\mathrm{e}_{k+i}$ represents an over-approximated reachable set for the state $ x_\mathrm{e}(k+i)$ of the error dynamics system~\eqref{Eq:ErrorDynamics}.
	\end{lemma}

\textit{Proof:} From the error dynamics system~\eqref{Eq:ErrorDynamics}, the error state reachable set can be computed using the model:    
	\begin{equation}
	\mathcal{R}_{k+i+1}^\mathrm{e}=\begin{bmatrix}A~B~H~J\end{bmatrix} \left(\mathcal{R}_{k+i}^\mathrm{e} \times K{\mathcal{R}}_{k+i}^\mathrm{e} \times \mathcal{Z}_{\epsilon} \times \mathcal{Z}_{\vartheta} \right) +\mathcal{Z}_{\omega}.
	\end{equation}
Since $\begin{bmatrix}A~B~H~J\end{bmatrix} \in \mathcal{M}_{\scalebox{0.7}{ABHJ}}$, according to Lemma~\ref{Lemma:M_ABHJ}, and if $ \mathcal{R}_{k+i}^\mathrm{e} $ and $\hat{\mathcal{R}}_{k+i}^\mathrm{e}$ start from the same initial set, it is evident that ${\mathcal{R}}_{k+i+1}^\mathrm{e} \in \hat{\mathcal{R}}_{k+i+1}^\mathrm{e}$. Therefore, the recursive relation~\eqref{Eq:ReachableSet} holds, providing an over-approximated reachable set for the data-driven error state. $\hfill\blacksquare$

    \subsubsection{\method{RDeeP-LCC} Optimization Formulation}
    In this part, we proceed to design the \method{RDeeP-LCC} optimization formulation, which is extended from \method{DeeP-LCC}~\cite{wang2023deep}.

   \textit{a) Trajectory Definition:} At each time step $ k $, we define the state trajectory $ x_\mathrm{ini} $ over the past $ T_\mathrm{ini} $ steps and the future state trajectory $x_z$ of the nominal system  in the next $N$ steps as follows:
   \begin{equation}
   \label{Eq:StateTrajectory}
   \begin{cases}
   x_\mathrm{ini} &= \mathrm{col}(x(k-T_\mathrm{ini}),x(k-T_\mathrm{ini}+1),\ldots,x(k-1)), \\
   x_\mathrm{z} &= \mathrm{col}(x_\mathrm{z}(k),x_\mathrm{z}(k+1),\ldots,x_\mathrm{z}(k+N-1)).
   \end{cases}
   \end{equation}
   The control trajectories $u_\mathrm{ini}$ and $u_\mathrm{z}$, the disturbance trajectories $\epsilon_{\mathrm{ini}}$ and $\epsilon_\mathrm{z}$, and the attack trajectories $\vartheta_{\mathrm{ini}}$ and $\vartheta_\mathrm{z}$ in the past $T_\mathrm{ini}$ steps and future $ N $ steps are defined similarly as in~\eqref{Eq:StateTrajectory}.

    \textit{b) Cost Function:} Similarly to \method{DeeP-LCC}~\cite{wang2023deep}, we utilize the quadratic function $ J(x_\mathrm{z},u_\mathrm{z}) $ to quantify the control performance, defined as follows:
    \begin{equation}
    \label{Eq:Cost}
    J(x_\mathrm{z},u_\mathrm{z})=\sum_{i=0}^{N-1}\left(\|x_\mathrm{z}(k+i)\|_{Q}^{2}+\|u_\mathrm{z}(k+i)\|_{R}^{2}\right),
    \end{equation}	
    where $ Q = \operatorname{diag}\left(Q_x, \xi Q_x, \ldots, \xi^{(n-1)} Q_x\right) \in \mathbb{R}^{2n \times 2n}$ and $ R\in \mathbb{R}$ are the weight matrices penalizing the system states and control inputs, with $ 0 < \xi \le 1 $ denotes the decay factor.

     \textit{c) Constraints:} The safety of the mixed platoon system is ensured by imposing the following constraints: 
     \begin{equation}
     \label{Eq:Constrain}
     \begin{cases}
     x(k+i) \in \mathcal{X},  \\
     u(k+i) \in \mathcal{U},
     \end{cases}
     \end{equation}
     where $\mathcal{X}=\left\{x(k) \in \mathbb{R}^{2n}\mid|x(k)|\leq \mathbf{1}_{n} \otimes x_{\max }\right\}$ is the state constraint, with $x_{\max }=\left[\tilde{s}_{\max }, \tilde{v}_{\max }\right]^{\top} $, where $\tilde{s}_{\max }$ and $\tilde{v}_{\max }$ are the constraint limits for spacing deviation and velocity deviation, respectively. For the control input constraint, we define $\mathcal{U}=\left\{u(k) \in \mathbb{R}\mid|u(k)|\leq u_{\max }\right\}$, where $u_{\max }$ denotes the maximum control input for the CAVs.
     
     Combining~\eqref{Eq:ReachableSet} and~\eqref{Eq:Constrain}, the constraint set for the nominal system~\eqref{Eq:NominalDynamics} can be calculated as follows:
     \begin{equation}
     \label{Eq:RConstraint}
     \begin{cases}
     \mathcal{X}_\mathrm{z}(k+i)= \mathcal{X} - \hat{\mathcal{R}}_{k+i}^\mathrm{e},\\
     \mathcal{U}_\mathrm{z}(k+i)= \mathcal{U} - K\hat{\mathcal{R}}_{k+i}^\mathrm{e},
     \end{cases}
     \end{equation}
     which yields the constraints for the predicted trajectory of the nominal system~\eqref{Eq:NominalDynamics}, given by:
     \begin{equation}
     \label{Eq:NormalConstraint}
     \begin{cases}
     x_\mathrm{z}(k+i) \in \mathcal{X}_\mathrm{z}(k+i),\\
     u_\mathrm{z}(k+i) \in \mathcal{U}_\mathrm{z}(k+i).
     \end{cases}
     \end{equation}
     
     For the future disturbance sequence $\epsilon_\mathrm{z}$ and attack sequence $\vartheta_\mathrm{z}$ of the nominal system, recalling~\eqref{Eq:DisturbanceDecouple}, we have
     \begin{equation}
     \label{Eq:DisturbanceConstraint}
     \epsilon_\mathrm{z}(k+i) = 0,\quad \vartheta_\mathrm{z}(k+i) = 0.
     \end{equation}

     \textit{d) Data-Driven Dynamics:} Based on Willems' fundamental lemma in~\cite[Theorem 1]{willems2005note}, \cite[Proposition 2]{wang2023deep}, and~\cite[Lemma 4.2]{coulson2019data}, the data-driven dynamics of the mixed platoon system can be given by
    \begin{equation}
    \label{Eq:WilliemExpandingSlack}
    \begin{bmatrix}
    X_{\mathrm{p}} \\
    U_{\mathrm{p}} \\
    E_{\mathrm{p}} \\
    F_{\mathrm{p}} \\
    X_{\mathrm{f}} \\
    U_{\mathrm{f}} \\
    E_{\mathrm{f}} \\
    F_{\mathrm{f}} 
    \end{bmatrix} g=\begin{bmatrix}
    x_{\mathrm {ini}} \\
    u_{\mathrm {ini}} \\
    \epsilon_{\mathrm{ini}} \\
    \vartheta_{\mathrm{ini}} \\
    x_\mathrm{z} \\
    u_\mathrm{z} \\
    \epsilon_\mathrm{z} \\
    \vartheta_\mathrm{z}
    \end{bmatrix} +
    \begin{bmatrix}
    \sigma \\
    0\\
    0 \\
    0 \\
    0 \\
    0 \\
    0 \\
    0
    \end{bmatrix},
    \end{equation}
    where $ \sigma \in \mathbb{R}^{2nT_\mathrm{ini}} $ is a slack variable ensuring feasibility. The existence of $ g \in \mathbb{R}^{T-T_\mathrm{ini}-N+1} $ satisfying~\eqref{Eq:WilliemExpandingSlack} implies that $ x_\mathrm{z} $, $ u_\mathrm{z} $, $\epsilon_\mathrm{z}$, and $\vartheta_\mathrm{z}$ form a future trajectory of length $N$.

    \textit{\method{e) RDeeP-LCC} Optimization Problem:} Naturally, we could formulate the optimization problem to solve the control input for the CAVs in mixed platoon systems as follows:
    \begin{equation}
    \label{Eq:OptimizationProblemFinal}
    \begin{aligned}
    \min\limits_{g, u_\mathrm{z}, x_\mathrm{z},\sigma} \; &J(x_\mathrm{z}, u_\mathrm{z}) + \lambda_{g}\|g\|_{2}^{2}+\lambda_{\sigma}\|\sigma\|_{2}^{2} \\
    \mathrm{ s.t. } 
   \quad & \eqref{Eq:NormalConstraint},~\eqref{Eq:DisturbanceConstraint},~\eqref{Eq:WilliemExpandingSlack},
    \end{aligned}
    \end{equation}
    where $\lambda_{g}$ and $ \lambda_{\sigma} $ denote the regularization penalty coefficients for the weighted two-norm for $ g $ and $\sigma $, respectively.
    
    Solving~\eqref{Eq:OptimizationProblemFinal} yields an optimal control sequence $ u_\mathrm{z}$ and the predicted state sequence $x_\mathrm{z}$ of the nominal system. Then, by
    \begin{equation}
    \label{Eq:UFinal}
    u(k) = u_\mathrm{z}(k) + K({x}(k)-{x}_\mathrm{z}(k)),
    \end{equation}
    we obtain the control input for the CAV, where ${x}(k)$ is measured from the actual system, and $K$  from~\eqref{Eq:K}.

    For each time step $k$ of the online data-driven predictive control, we solve the final \method{RDeeP-LCC} optimization problem~\eqref{Eq:OptimizationProblemFinal} using the receding horizon technique. The detailed procedure of \method{RDeeP-LCC} is presented in Algorithm~\ref{algorithm1}.

    \begin{algorithm}[t]
    \caption{\method{RDeeP-LCC}}
    \label{algorithm1}
    \SetAlgoLined
    \KwIn{Collected data ($U$, $E$, $F$, $X$), constraints ($\mathcal{X}$, $\mathcal{U}$), weight ($Q$, $R$), bounded $\mathcal{Z}_{{\epsilon}}$, $\mathcal{Z}_{\vartheta}$, and $\mathcal{Z}_{\omega}$, past horizon $T_\mathrm{ini}$, control horizon $N$, total number of steps $N_\mathrm{f}$.}
    	
    Construct data ($U_-$, $E_-$, $F_-$, $X_-$, $X_+$)\;
    Offline construct $ U_{\mathrm{p}} $, $ U_{\mathrm{f}} $, $ E_{\mathrm{p}} $, $ E_{\mathrm{f}} $, $ F_{\mathrm{p}} $, $ F_{\mathrm{f}} $, $ X_{\mathrm{p}} $, $ X_{\mathrm{f}} $ using~\eqref{Eq:Hankel}, $\mathcal{M}_{\scalebox{0.7}{ABHJ}}$ using~\eqref{Eq:M_ABHJ}, and $K$ using~\eqref{Eq:K} \;
    Initialize past data ($x_{\mathrm {ini}}$, $u_{\mathrm {ini}}$, ${\epsilon}_\mathrm{ini}$, ${\vartheta}_\mathrm{ini}$) at the initial time step $0$\;

    \While{$ 0 \leq k \leq N_\mathrm{f} $}{
    	Compute reachable set $\hat{\mathcal{R}}_{k+i}^\mathrm{e}$ using~\eqref{Eq:ReachableSet}\;
    	Compute $\mathcal{X}_\mathrm{z}(k+i)$ and $\mathcal{U}_\mathrm{z}(k+i)$ using~\eqref{Eq:RConstraint}\;
    	Solve~\eqref{Eq:OptimizationProblemFinal} to obtain $u_\mathrm{z}$ and $x_\mathrm{z}$ for the nominal system\; 

        Measure the $x(k)$ from the actual system\;
            
        Obtain $ u(k) $ from~\eqref{Eq:UFinal} and apply it to the CAV\;
        $ k \gets k+1 $ and update ($x_{\mathrm {ini }}$, $u_{\mathrm {ini }}$, ${\epsilon}_\mathrm{ini}$, ${\vartheta}_\mathrm{ini}$)\;
    		
    	}
    \end{algorithm}

    \begin{remark}
    	It is worth noting that in~\eqref{Eq:OptimizationProblemFinal}, $x_{\mathrm {ini}},u_{\mathrm {ini}},\epsilon_{\mathrm {ini}},\vartheta_{\mathrm {ini}}$ represent the past trajectories of the actual system~\eqref{Eq:DiscreteSystem}, while $x_z, u_\mathrm{z},\epsilon_\mathrm{z},\vartheta_\mathrm{z}$ denote the predicted trajectories of the nominal system~\eqref{Eq:NominalDynamics}. As shown in~\eqref{Eq:DisturbanceDecouple}, we assume $\epsilon_\mathrm{z}(k) = 0$ and $\vartheta_\mathrm{z}(k) = 0$, and capture the actual disturbances $\epsilon_\mathrm{e}(k) = \epsilon(k)$ and actual attacks $\vartheta_\mathrm{e}(k) = \vartheta(k)$ solely in the error system~\eqref{Eq:ErrorDynamics}. Provided $\epsilon(k) \in \mathcal{Z}_{\epsilon}$ and $\vartheta(k) \in \mathcal{Z}_{\vartheta}$ in~\eqref{Eq:W_Zonotope}, the effects of disturbances and attacks are further incorporated into the calculation of the error reachable set~\eqref{Eq:ReachableSet}, resulting in a more stringent constraint~\eqref{Eq:NormalConstraint} on $x_\mathrm{z}$ and $u_\mathrm{z}$ of the nominal system. This approach addresses the influence of unknown future disturbances and attacks, which are often oversimplified as zero in previous research~\cite{wang2023deep,zhang2023privacy,li2024physics}.
    \end{remark}

       \begin{remark}
       Unlike the ZPC method in~\cite{lan2021data}, which relies on forward reachable sets for prediction and control input computation, the proposed \method{RDeeP-LCC} framework adopts a decomposition strategy in which the control input is separated into an error component and a nominal component. The error control input provides strict stability guarantees and is explicitly incorporated into the computation of the error reachable set, thereby reducing conservatism in calculating the feasible region. The nominal control input is derived by solving optimization problem~\eqref{Eq:OptimizationProblemFinal}, which leverages the data-driven dynamics constructed in~\eqref{Eq:WilliemExpandingSlack} from both offline and online data. The integration of online data enhances the adaptability of the framework to diverse and time-varying operating conditions.
    \end{remark}

    \section{Human-in-the-Loop Experiment}
    \label{Sec:5}
    In this section, we establish a human-in-the-loop bench experimental platform to verify the effectiveness of the proposed \method{RDeeP-LCC} method.

    \subsection{Human-in-the-Loop Experimental Platform}
\begin{figure}[t]
    	\centering
            \subfigure[Experimental devices]{
    	\includegraphics[width=8.5cm]{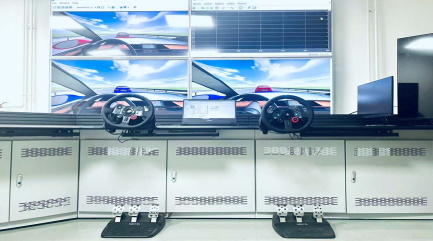}}
            \subfigure[Connection structure]{
    	\includegraphics[width=8.7cm]{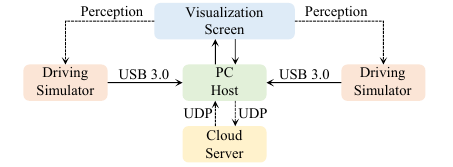}}
    	\vspace{-0.3cm}
    	\caption{Human-in-the-loop experimental platform for mixed platoon control.}
    	\label{Fig:ExperimentPlatform}
    \end{figure}

    To validate the effectiveness of the \method{RDeeP-LCC} method under noise and attacks, we developed a human-in-the-loop platform that replicates a real-world driving scenario using PreScan 8.5 software on a Windows 10 system. As depicted in~\figurename~\ref{Fig:ExperimentPlatform}, the setup comprises a visualization screen module, two Logitech G29 driving simulator modules (including brake pedal, accelerator pedal, and steering wheel), a PC host module equipped with an Intel Core i9-13900KF CPU and 64 GB of RAM, and a cloud server powered by an Intel(R) Xeon(R) Gold 6226R CPU @ 2.90GHz.
    
    In this setup, the experiment is conducted on a straight road. Drivers observe the preceding vehicles through a high-definition visualization screen module and control their vehicles using the Logitech G29 driving simulators. These simulators are connected to the PC host module's Simulink software via the USB 3.0, enabling the real-time transmission of control commands from the drivers to Simulink, including braking and acceleration maneuvers. Subsequently, Simulink relays these commands to PreScan 8.5 software, which displays the vehicle dynamic in real-time on the visualization screen. The detailed dynamics model of the simulated vehicles is based on the Audi A8 model provided by PreScan 8.5. The control algorithm is developed within the MATLAB/Simulink environment and deployed on a cloud server. Vehicle states and control inputs are transmitted between the PC host and the cloud server using UDP communication. Additionally, to simulate realistic driving conditions, we use the \textit{uniform random number} module in Simulink to simulate noise and attacks. The entire experiment operates at a frequency of $\SI{20}{Hz}$.

	\subsection{Experimental Setup}
    In our experiments, the \method{RDeeP-LCC} method or other baseline methods control the CAV (indexed as 1), while two drivers operate the HDVs (indexed as 2 and 3) using the Logitech G29 simulators. To ensure accuracy and consistency, before the formal experiment, drivers undergo a $ 3 $-hour training session in car-following scenarios to familiarize themselves with the driving simulators. Furthermore, to ensure fairness in the comparative experiment, all drivers are informed that the study involves mixed platoon control strategies. However, they are not informed of which specific control algorithm is implemented on the CAV, nor are they made aware of the evaluation indices used in the analysis. Throughout the experiment, each driver is instructed to maintain their natural car-following behavior. To replicate a realistic traffic scenario, the HV (indexed as $0$) is assigned a time-varying velocity profile derived from the SFTP-US06 driving cycle, while subject to a noise boundary of $\omega_\mathrm{max} = 0.02$ and a state-independent attack boundary of $\vartheta_\mathrm{max} = 2$.

\begin{figure}[t]
    	\centering
            \vspace{-2mm}
            \subfigure[Fitting results for HDV $2$]{
    	\includegraphics[width=4.2cm]{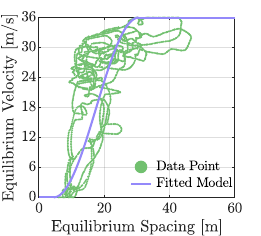}}
            \subfigure[Fitting results for HDV $3$]{
    	\includegraphics[width=4.2cm]{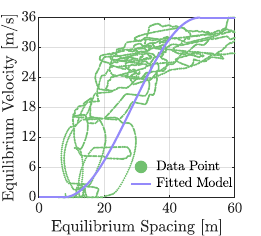}}
    	\vspace{-0.2cm}
    	\caption{The fitting results of equilibrium spacings and equilibrium velocities for the two HDVs.}
    	\label{Fig:HDVsFitting}
    \end{figure}

    In this experiment, we initially set the velocity of the HV as the equilibrium velocity $v^{*}_i$. However, it is a challenge for us to get the equilibrium spacing $ s^{*}_i $ for real drives. To address this issue, we conduct a preliminary experiment where two drivers operate their vehicles in a car-following scenario using the SFTP-US06 driving cycle. Subsequently, we gather experimental data, represented by the green points in~\figurename~\ref{Fig:HDVsFitting}. This data is then utilized to fit the OVM model. The fitting results for two drivers, namely HDV $2$ and HDV $3$, are illustrated by the purple lines in~\figurename~\ref{Fig:HDVsFitting}(a) and (b), respectively. The parameters obtained from the fitting process for the two drivers are as follows: for HDV $2$, $v_{\max } =\SI{36}{m/s}$, $s_{\min } = \SI{4.6}{m}$, $s_{\max }=\SI{30.6}{m}$; for HDV $3$, $v_{\max } =\SI{36}{m/s}$, $s_{\min } = \SI{7.5}{m}$, $s_{\max }=\SI{49.4}{m}$. By employing the fitted model, we are able to make estimations regarding the equilibrium spacing $ s^{*}_i $ in the online predictive control phase.

In the formal experiments for method validation, we first conduct data collection of the mixed platoons. Around the equilibrium velocity of $ v^{*}=\SI{18}{m/s} $, we generate random control inputs of the CAV by $u(t) \sim \mathbb{U}\left[-0.2, 0.2\right] $, random disturbance inputs by  $\epsilon(t) \sim \mathbb{U}\left[-0.5, 0.5\right] $, and random attack inputs by $\vartheta(t) \sim \mathbb{U}\left[-0.3, 0.3\right] $, where $\mathbb{U}$ represents uniform distribution. The offline collected trajectories, with a length of $ T = 600 $ and a sampling interval of $\SI{0.05}{s}$, are then employed to construct~\eqref{Eq:data_sequence}. In the offline phase, based on the collected data, we proceed with the construction of the matrix zonotope set $\mathcal{M}_{\scalebox{0.7}{ABHJ}}$ in~\eqref{Eq:M_ABHJ}, the feedback control law $K$ in~\eqref{Eq:K}, and the Hankel matrices in~\eqref{Eq:Hankel}. In the online predictive control phase, the $ N=5 $ and $ T_\mathrm{ini}= 20 $ for the state trajectory~\eqref{Eq:StateTrajectory}. The cost~\eqref{Eq:Cost} is configured with $ \xi=0.6 $, $Q_x = \operatorname{diag}\left(0.5, 1\right)$, and $R = 0.1$. Constraints are imposed as $x_{\max} = \left[7, 7\right]^{\top}$ and $u_{\max} = 5$ in~\eqref{Eq:Constrain}. In the optimization formulation~\eqref{Eq:OptimizationProblemFinal}, we use $\lambda_{g} = 10$ and $\lambda_{\sigma} = 10$. The step length is $ t_\mathrm{s} = \SI{0.05}{s} $. The HV initiates its motion according to the predefined velocity, while the CAV and two HDVs respond to the controller's commands and the drivers' actions, respectively.

For comparison, several baseline methods are considered: 1) the standard MPC method, which assumes full knowledge of system dynamics in~\eqref{Eq:DiscreteSystem}; 2) the standard \method{DeeP-LCC} method from~\cite{wang2023deep}; and 3) the ZPC method from~\cite{lan2021data}, explicitly accounting for noise, disturbances, and attacks. All baseline methods share the same parameter values as \method{RDeeP-LCC}, except that standard MPC and standard \method{DeeP-LCC} use $N = 10$. Additionally, the standard \method{DeeP-LCC} and ZPC utilize the identical data sets with \method{RDeeP-LCC}. Furthermore, we also include an all-HDV scenario as a baseline, where the CAV is controlled using the OVM model. For ZPC and \method{RDeeP-LCC}, reachable sets are computed using the CORA 2021 toolbox~\cite{althoff2015introduction}. To enhance computational efficiency, we employ the interval set to over-approximate the zonotope set. The \textit{interval} command is detailed in the CORA 2021 Manual~\cite{althoff2015introduction}.

To quantitatively evaluate the performance of the proposed and baseline control methods, multiple performance indices are introduced. First, the velocity mean absolute deviation $R_\mathrm{v}$ is used to assess tracking accuracy, defined as
\begin{equation}
\label{Eq:IndexV}
R_\mathrm{v} = \frac{1}{(t_\mathrm{f}-t_\mathrm{0})}\frac{1}{n} \sum_{k=t_\mathrm{0}}^{t_\mathrm{f}} \sum_{i=1}^{n} \left|v_{i}(k) - v^{*}(k)\right|,
\end{equation}
where $t_\mathrm{0}$ and $t_\mathrm{f}$ denote the initial and final time steps, respectively.

Next, the control performance is quantified using the actual cost value $R_\mathrm{c}$, computed according to~\eqref{Eq:Cost}:
\begin{equation}
\label{Eq:IndexCost}
R_\mathrm{c} = \sum_{k=t_\mathrm{0}}^{t_\mathrm{f}} \left(\|x(k)\|_{Q}^{2} + \|u(k)\|_{R}^{2}\right),
\end{equation}

In addition, fuel consumption and driving comfort are evaluated through $R_\mathrm{f}$ and $R_\mathrm{a}$, defined as
\begin{equation}
\label{Eq:IndexFuel}
R_\mathrm{f} = t_\mathrm{s} \sum_{k=t_\mathrm{0}}^{t_\mathrm{f}} \sum_{i=1}^{n} f_i,
\end{equation}
\begin{equation}
\label{Eq:IndexComfort}
R_\mathrm{a} = \frac{1}{(t_\mathrm{f}-t_\mathrm{0})}\frac{1}{n} \sum_{k=t_\mathrm{0}}^{t_\mathrm{f}} \sum_{i=1}^{n} a_i^2,
\end{equation}
where $f_i$ (\SI{}{mL/s}) denotes the fuel consumption rate of vehicle $i$, calculated using the model in~\cite{wang2023deep}, and $a_i$ is its acceleration.

Finally, to evaluate constraint satisfaction and safety performance, we count the number of violations of the safety constraint in~\eqref{Eq:Constrain} under each control strategy, denoted as $R_\mathrm{n}$.

Together, these indices comprehensively capture the control performance, energy efficiency, ride comfort, and safety characteristics of the proposed approach compared with baseline methods.

    \subsection{Experiment 1: State-Independent Attacks}
    \label{Sec:5-D}
\begin{figure*}[t]
   	\centering
   	\subfigure[Standard MPC]{
   		\includegraphics[width=8.75cm]{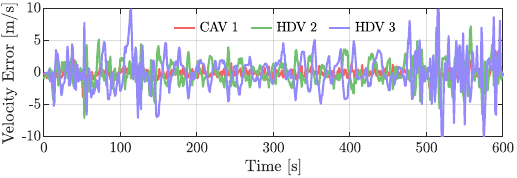}}
   	\vspace{-0.1cm}
   	\subfigure[Standard \method{DeeP-LCC}]{
   		\includegraphics[width=8.75cm]{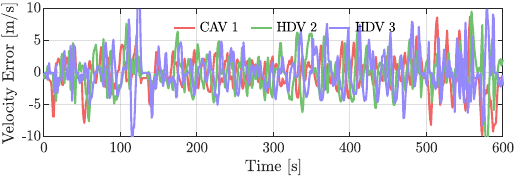}}\\
   	\vspace{-0.1cm}
   	\subfigure[ZPC]{
   		\includegraphics[width=8.75cm]{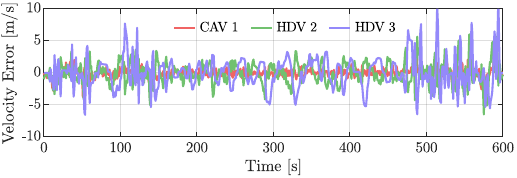}}
        \vspace{-0.1cm}
   	\subfigure[\method{RDeeP-LCC}]{
   		\includegraphics[width=8.75cm]{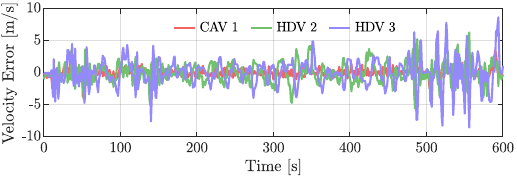}}\\
   	\vspace{-0.1cm}
   	\caption{Velocity errors in mixed platoon human-in-the-loop experiments for five control methods under state-independent attacks. The red profiles represent the velocity error between the HV and the CAV, the green profiles show the velocity error between the CAV and the HDV 2, and the purple profiles depict the velocity error between the HDV 2 and the HDV 3.}
   	\label{Fig:ExperimentResult}
    \end{figure*}

    The experimental results are presented in~\figurename~\ref{Fig:ExperimentResult}. In the presence of noise and attacks, significant velocity errors between adjacent vehicles are observed in the standard \method{DeeP-LCC}, as shown in~\figurename~\ref{Fig:ExperimentResult}(b), indicating poor tracking capability in these scenarios. This performance degradation is attributed to the heavy reliance of the standard \method{DeeP-LCC} on data for modeling~\eqref{Eq:WilliemExpandingSlack}, where the presence of data noise significantly reduces model accuracy. Additionally, standard \method{DeeP-LCC} lacks preemptive robustness against attacks, further compromising its control performance in environments with coexisting noise and attacks. In contrast, standard MPC, ZPC, and \method{RDeeP-LCC} significantly improve the tracking performance of the mixed platoon. Furthermore, a comparison between~\figurename~\ref{Fig:ExperimentResult}(a), \figurename~\ref{Fig:ExperimentResult}(c), and~\figurename~\ref{Fig:ExperimentResult}(d) reveals that the vehicles' velocity error in the \method{RDeeP-LCC} method is even smaller. This outcome underscores the exceptional tracking performance of the \method{RDeeP-LCC} method, highlighting its robustness and efficacy in achieving precise control under diverse traffic conditions.

   Table~\ref{tab:experiment_B} provides the performance of all controllers across five evaluation indices ($R_\mathrm{v}$, $R_\mathrm{c}$, $R_\mathrm{f}$, $R_\mathrm{a}$, $R_\mathrm{n}$) under state-independent attacks. Compared with the all HDVs, the proposed \method{RDeeP-LCC} achieves improvements of $26.1\%$, $24.7\%$, $11.4\%$, $26.0\%$, and $93.5\%$ across the respective indices. Although the standard MPC and ZPC methods achieve notable improvements over all HDVs, their performance remains inferior to that of our method. Notably, the standard \method{DeeP-LCC} method reports the worst performance for all indices. This outcome suggests that purely data-driven predictive control, without robust design mechanisms, may struggle to meet desired performance standards in the presence of noise and attacks. In fact, such approaches might even exhibit worse performance compared to the inherent capabilities of all HDVs. In contrast, our \method{RDeeP-LCC} approach focuses on optimizing constraints using reachable set analysis, seeking to ensure that the state of a mixed platoon system remains within a tighter set of safety constraints, thereby reducing $R_\mathrm{n}$. Moreover, the proposed \method{RDeeP-LCC} consistently outperforms ZPC across all indices, primarily due to its robust tube-based control design in~\eqref{Eq:DynamicsDecomposition} and the adaptive data-driven dynamics in~\eqref{Eq:WilliemExpandingSlack}, which leverage online data for real-time updates. Together, these features enable a well-balanced trade-off between robustness and conservatism, ensuring superior performance in dynamic environments. Overall, the proposed \method{RDeeP-LCC} framework improves tracking accuracy, control efficiency, energy economy, driving comfort, and overall safety in mixed platoon control systems under noise and attacks.

\begin{table*}[t]
\centering
\renewcommand{\arraystretch}{1.6}
\caption{Performance Indices in Human-in-the-loop Experiment under State-Independent Attacks}
\label{tab:experiment_B}
\begin{tabular}{m{1.6cm}<{\centering} m{2.5cm}<{\centering} m{3.0cm}<{\centering} m{2.5cm}<{\centering} m{2.5cm}<{\centering} m{2.5cm}<{\centering}}
\hline
Index & $R_\mathrm{v} [\SI{}{m/s}]$ & $R_\mathrm{c} [\SI{}{-}]$ & $R_\mathrm{f} [\SI{}{mL}]$ & $R_\mathrm{a} [\SI{}{m^2/s^4}]$ & $R_\mathrm{n} [\SI{}{-}]$\\ \hline
All HDVs & $1.65$ $(----)$ & $0.97 \times 10^{6}$ $(----)$ & $8133$ $(----)$ & $3.88$ $(----)$ & $990$ $(----)$\\ \hline
MPC & $1.34$ $(\downarrow 18.8\%)$ & $0.81 \times 10^{6}$ $(\downarrow 16.5\%)$ & $7687$ $(\downarrow 5.5\%)$ & $3.66$ $(\downarrow 5.7\%)$ & $219$ $(\downarrow 77.9\%)$\\ \hline
\method{DeeP-LCC} & $2.56$ $(\uparrow 55.2\%)$ & $3.26 \times 10^{6}$ $(\uparrow 236.1\%)$ & $8404$ $(\uparrow 3.3\%)$ & $3.98$ $(\uparrow 2.6\%)$ & $1535$ $(\uparrow 55.1\%)$\\ \hline
ZPC & $1.25$ $(\downarrow 24.2\%)$ & $0.75 \times 10^{6}$ $(\downarrow 22.7\%)$ & $7341$ $(\downarrow 9.7\%)$ & $2.93$ $(\downarrow 24.5\%)$ & $77$ $(\downarrow 92.2\%)$\\ \hline
\method{RDeeP-LCC} & $\mathbf{1.22}$ $(\downarrow \mathbf{26.1\%})$ & $\mathbf{0.73 \times 10^{6}}$ $(\downarrow \mathbf{24.7\%})$ & $\mathbf{7205}$ $(\downarrow \mathbf{11.4\%})$ & $\mathbf{2.87}$ $(\downarrow \mathbf{26.0\%})$ & $\mathbf{64}$ $(\downarrow \mathbf{93.5\%})$\\ \hline
\end{tabular}
\vspace{-3mm}
\end{table*}

To evaluate the computational efficiency of the proposed method, we analyze its offline and online phases. In the offline phase, constructing the over-approximated system matrix set $\mathcal{M}_{\scalebox{0.7}{ABHJ}}$ in~\eqref{Eq:M_ABHJ} requires $\SI{0.15}{s}$, computing the data-driven stabilizing feedback control law $K$ in~\eqref{Eq:K} takes $\SI{0.65}{s}$, and generating the Hankel matrices in~\eqref{Eq:Hankel} is completed within $\SI{0.03}{s}$. Notably, $\mathcal{M}{\scalebox{0.7}{ABHJ}}$ is also used in ZPC, and the Hankel matrices are shared with \method{DeeP-LCC}, ensuring consistency across these methods. In the online phase, the standard MPC exhibits the fastest execution time of $\SI{0.001}{s}$. The ZPC requires a modest increase to $\SI{0.018}{s}$ due to the online computation of reachable sets. The \method{DeeP-LCC} requires $\SI{0.024}{s}$, mainly attributed to solving an optimization problem with decision variable $g$ of size $T - T_\mathrm{ini} - N + 1$. Similarly, the proposed \method{RDeeP-LCC} takes $\SI{0.038}{s}$. Notably, the online computation time of \method{RDeeP-LCC} remains within the $\SI{0.05}{s}$, ensuring its real-time applicability.

\subsection{Experiment 2: State-Dependent Attacks}
\label{Sec:5-E}
To further assess the robustness of the proposed control framework under adversarial conditions, experiments are conducted with state-dependent attacks. Unlike the bounded random signals used in the state-independent attack scenario in Section~\ref{Sec:5-D}, the state-dependent attack is formulated following~\cite{zhang2023adaptive} as $\vartheta(k)=\tilde{v}_1^3(k)\cos(\tilde{v}_1(k))+2\sin(\tilde{v}_1(k))$.  All other experimental configurations are kept identical to those described in Section~\ref{Sec:5-D} to ensure consistency in performance comparison.

\begin{figure*}[t]
   	\centering
   	\subfigure[Standard MPC]{
   		\includegraphics[width=8.75cm]{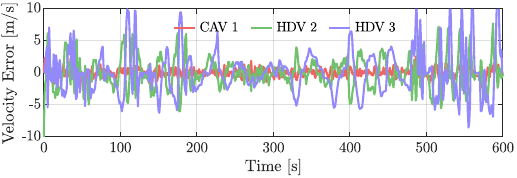}}
   	\vspace{-0.1cm}
   	\subfigure[Standard \method{DeeP-LCC}]{
   		\includegraphics[width=8.75cm]{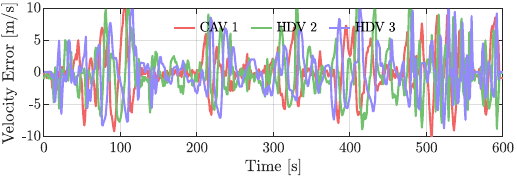}}\\
   	\vspace{-0.1cm}
   	\subfigure[ZPC]{
   		\includegraphics[width=8.75cm]{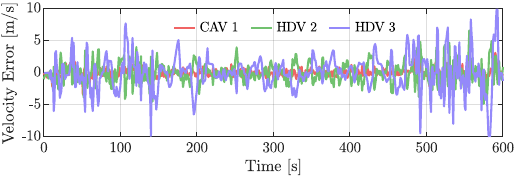}}
        \vspace{-0.1cm}
   	\subfigure[\method{RDeeP-LCC}]{
   		\includegraphics[width=8.75cm]{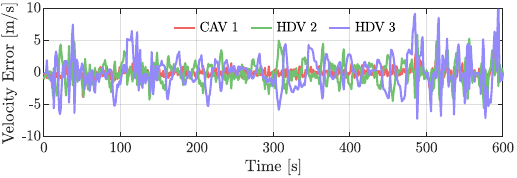}}\\
   	\vspace{-0.1cm}
   	\caption{Velocity errors in mixed platoon human-in-the-loop experiments for five control methods under state-dependent attacks. The red profiles represent the velocity error between the HV and the CAV, the green profiles show the velocity error between the CAV and the HDV 2, and the purple profiles depict the velocity error between the HDV 2 and the HDV 3.}
   	\label{Fig:ExperimentResult_2}
    \end{figure*}

The results of these experiments are illustrated in Fig.~\ref{Fig:ExperimentResult_2}. It can be observed that the proposed \method{RDeeP-LCC} consistently achieves the lowest velocity tracking error, particularly improving upon the baseline \method{DeeP-LCC} method. This improvement demonstrates that the integration of reachability analysis effectively enhances robustness, addressing the limitations of \method{DeeP-LCC} under adversarial conditions. The quantitative evaluation of performance indices under state-dependent attacks is summarized in Table~\ref{tab:experiment_B_Sin}. Relative to the all HDVs, \method{RDeeP-LCC} achieves improvements of $25.4\%$, $22.1\%$, $12.9\%$, $32.5\%$, and $93.8\%$ across the respective indices, reflecting substantial gains in tracking accuracy, control efficiency, energy consumption, driving comfort, and driving safety. These results confirm that the proposed \method{RDeeP-LCC} framework consistently delivers superior performance under diverse attack types, validating the effectiveness and robustness of its design.

\begin{table*}[t]
	\centering
	\renewcommand{\arraystretch}{1.6}
	\caption{Performance Indices in Human-in-the-loop Experiment under State-Dependent Attacks}
	\label{tab:experiment_B_Sin}
    \begin{tabular}{m{1.6cm}<{\centering} m{2.5cm}<{\centering} m{3.0cm}<{\centering} m{2.5cm}<{\centering} m{2.5cm}<{\centering} m{2.5cm}<{\centering}}
    \hline
    Index & $R_\mathrm{v} [\SI{}{m/s}]$ & $R_\mathrm{c} [\SI{}{-}]$ & $R_\mathrm{f} [\SI{}{mL}]$ & $R_\mathrm{a} [\SI{}{m^2/s^4}]$ & $R_\mathrm{n} [\SI{}{-}]$\\ \hline
    All HDVs & $1.85$ $(----)$ & $1.13 \times 10^{6}$ $(----)$ & $8350$ $(----)$ & $4.31$ $(----)$ & $1125$ $(----)$\\ \hline
    MPC & $1.61$ $(\downarrow 13.0\%)$ & $0.99 \times 10^{6}$ $(\downarrow 12.4\%)$ & $8079$ $(\downarrow 3.2\%)$ & $4.01$ $(\downarrow 7.0\%)$ & $317$ $(\downarrow 71.8\%)$\\ \hline
    \method{DeeP-LCC} & $2.51$ $(\uparrow 35.7\%)$ & $3.72 \times 10^{6}$ $(\uparrow 229.2\%)$ & $8523$ $(\uparrow 2.1\%)$ & $4.54$ $(\uparrow 5.3\%)$ & $1870$ $(\uparrow 66.2\%)$\\ \hline
    ZPC & $1.43$ $(\downarrow 22.7\%)$ & $0.89 \times 10^{6}$ $(\downarrow 21.2\%)$ & $7422$ $(\downarrow 11.1\%)$ & $3.18$ $(\downarrow 26.2\%)$ & $82$ $(\downarrow 92.7\%)$\\ \hline
    \method{RDeeP-LCC} & $\mathbf{1.38}$ $(\downarrow \mathbf{25.4\%})$ & $\mathbf{0.88 \times 10^{6}}$ $(\downarrow \mathbf{22.1\%})$ & $\mathbf{7272}$ $(\downarrow \mathbf{12.9\%})$ & $\mathbf{2.91}$ $(\downarrow \mathbf{32.5\%})$ & $\mathbf{70}$ $(\downarrow \mathbf{93.8\%})$\\ \hline
    \end{tabular}
	\vspace{-3mm}
\end{table*}

	\section{Conclusion}
	\label{Sec:6}
	In this paper, we propose a novel \method{RDeeP-LCC} method for mixed platoon control under conditions of data noise and adversarial attacks. The \method{RDeeP-LCC} method incorporates the unknown dynamics for HDVs and directly relies on trajectory data of mixed platoons to construct data-driven reachable set constraints and a data-driven predictive controller, which provides safe and robust control inputs for CAVs. To validate the effectiveness and superiority of this method, human-in-the-loop experiments are conducted. The results indicate that the \method{RDeeP-LCC} method has significantly improved the tracking performance of mixed platoons in the presence of data noise and adversarial attacks.
	
	In future research, one practical concern in \method{RDeeP-LCC} is the influence of communication delays induced by actuators and communication. Another interesting topic is to improve the real-time computational efficiency of data-driven predictive control methods, which facilitates scalable deployment in real-world traffic scenarios. Furthermore, it is promising to develop nonlinear versions of the \method{RDeeP-LCC} method, which can reduce conservatism and further improve the control performance of mixed platoon systems.


    \bibliographystyle{IEEEtran}
    \bibliography{IEEEabrv,Reference}

\vfill
	
\end{document}